\documentclass[superscriptaddress,amsmath,amssymb,floatfix,preprint]{revtex4-1}

\usepackage{amsmath}
\usepackage{amssymb}
\usepackage{amsthm}
\usepackage{mathrsfs}
\usepackage{graphicx}
\usepackage{caption}
\usepackage{subcaption}

\usepackage[pagebackref]{hyperref}

\numberwithin{equation}{section}

\graphicspath{{./Figures/}}

\usepackage[letterpaper]{geometry}

\theoremstyle{definition}

\newcommand{\ve}{\varepsilon}

\newcommand{\N}{\mathbb{N}}

\begin{document}

\title{Precision studies of $v_{n}$ fluctuations}

\author{Tyler Gorda}
\author{Paul Romatschke}
\affiliation{University of Colorado Boulder, Boulder, CO}

\date{\today}

\begin{abstract}
	The power spectrum of heavy ion collisions is investigated by studying initial state fluctuations
	on top of a smooth hydrodynamic flow.  In particular, the stability
	of the location of the first minimum of the power spectrum and the dependence of the hydrodynamic response $v_{n} / \ve_{n}$ on $n$ and on
	$\eta / s$ are discussed. In our study we develop a new Green's function method for the analytic hydrodynamic flow by S. Gubser and make use 
	of a fully non-linear hydrodynamics code.
	We find that there will be no well-defined first minimum of the response for $n < 10$, due to the fact that all minima in that region are
	found to be sensitive to the location of the initial perturbations. Also, we find that the often proposed form of the hydrodynamical
	response, that $\ln \left( v_{n} / \ve_{n} \right)$ depend quadratically on $n$ and linearly on $\eta / s$, 
	should not hold once many events have been averaged over.
\end{abstract}

\maketitle

\section{Introduction}
\label{sec: intro}

	Experiments at the Relativistic Heavy Ion Collider (RHIC) and the Large Hadron Collider (LHC) have both been able to produce and detect 
	an exotic phase of matter, the quark-gluon plasma (QGP), by colliding massive nuclei at relativistic energies.  The energy densities
	involved in these collisions are so intense that the hadrons participating in the collisions dissolve into their constituent partons,
	creating a state of matter in which the color charge is no longer confined.  Despite the short lifetime of this matter (on the order of
	$10$ fm/c), there is strong evidence that it has time to (locally) thermalize and can be described as a nearly perfect fluid.  One can
	compute the expected abundance of different particle species produced from a thermalized QGP and compare it with the
	ratios of detected particle species at RHIC or LHC; the results are in good agreement with one another \cite{Sorensen}, \cite{Thermal QGP}.
	Another place where hydrodynamics has provided particularly good agreement with heavy ion experiments is with the so-called "collective flow" 
	\cite{Sorensen} \cite{Shuryak}. When two nuclei collide the resulting QGP will be anisotropic in space, because of differences in density 
	distributions inside the individual nuclei, even for central collisions.  Within hydrodynamics (implying strong interactions), pressure 
	gradients will convert this spatial anisotropy into an anisotropy in momentum space.  This effect would not be observed if the matter 
	produced were noninteracting. 

	The full power spectrum resulting from heavy ion collisions is given by the Fourier Transform of the two particle correlation function 
	\cite{vn paper} 
\begin{equation}
		\frac{d N}{d \Delta \phi} = \left\langle \frac{d N}{d \Delta \phi} \right\rangle 
			\left( 1 + 2 \sum_{m = 1}^{\infty} |v_{m}|^{2} \cos (m \Delta \phi) \right) .
\end{equation}	
	Here $N$ is the particle count and $\Delta \phi$ is the relative momentum angle between two observed particles.
	The elliptic flow is controlled by the coefficient $v_{2}$.  From this power spectrum, one should be able to extract information about
	the initial density fluctuations or other physics in the QGP, in analogy with the Cosmic Microwave Background power spectrum.	

	One way to theoretically investigate this power spectrum is by studying hydrodynamic fluctuations on top of a smooth fluid flow and
	following how these perturbations propagate through to the particle spectra and on to the $v_{m}$.  This approach was taken by Staig and
	Shuryak in a series of research papers \cite{Shuryak} \cite{Staig 1} \cite{Staig 2} studying linear perturbations.  One can also study
	the power spectrum numerically using fully nonlinear hydrodynamic codes. 

	Both of these approaches require much analytic or numerical effort to obtain the power spectrum.  In order to obviate the need for this
	extensive effort every time one wants to compare theory and experiment, focus has been placed on determining how particular features of 
	the power spectrum are affected by properties of the QGP itself, e.g. the viscosity or freezeout temperature.  Particular interest 
	has been placed on the first minimum of the power spectrum \cite{Staig 2} or on the hydrodynamic response of the system $v_{m} / e_{m}$
	where the $e_{m}$, defined by
\begin{equation}
		e_{m} = \left| \frac{\int \! r^{m} \ve( x ) e^{i m \phi} d^{2} x}{\int \! r^{m} \ve( x ) d^{2} x} \right|, 
\end{equation}
	are the initial state eccentricities. Here $\ve \propto T^{4}$ is the initial energy density, and the integrals are over the transverse 
	plane of the collision. This response is often proposed to depend on hydrodynamic parameters by 
\begin{equation}
		\ln \left( \frac{v_{m}}{e_{m}} \right) \propto - \frac{4}{3 R T} m^{2} \frac{\eta}{s},
\end{equation}
	where $R$ is the transverse size of the collision, $T$ is the temperature of the collision, $\eta$ is the shear viscosity, and $s$ is the
	entropy density.  This dependence of the response has been investigated by Lacey \emph{et al.} \cite{Lacey}.  One of the questions we 
	investigate in this paper is how robust these conclusions are; that is, can information about 
	the initial state fluctuations or about properties of the QGP actually be extracted once one has averaged over many events?  

\subsection{Analytically Known Flows: Bjorken and Gubser Flow}
\label{subsec: Gubser Review}

	Full, nonlinear hydrodynamic flow generally requires computer code in order to generate
	results, because the full equations are too complicated to solve analytically. 
	However, one can often investigate the approximate behavior of the full solutions by considering small perturbations around a known
	analytical flow. If one possesses a given analytical flow, i.e. its energy-momentum tensor $T^{\mu \nu}_{0}$, then one can consider
	energy momentum tensors of the form
\begin{equation}
		T^{\mu \nu} = T^{\mu \nu}_{0} + \delta T^{\mu \nu},
\end{equation}
	where $\delta T^{\mu \nu}$ is small to the first order.  Taking the covariant derivative of this perturbed energy-momentum tensor leads
	directly to the first order equations for the perturbation.  In what follows we will consider perturbations to the energy
	density, pressure, and four velocity directly, which make up the energy momentum tensor
\begin{equation}
		T^{\mu \nu} = (\ve + p) u^{\mu} u^{\nu} + p \, g^{\mu \nu} - 2 \eta \, \sigma^{\mu \nu} - \zeta \, ( \nabla_{\gamma} 
			u^{\gamma} ) \Delta^{\mu \nu}
\end{equation}
	where $\ve$ is the energy density, $p$ is the pressure, $u^{\mu}$ is the four-velocity, $g_{\mu \nu}$ is the metric, $\eta$ is the shear 
	viscosity, $\zeta$ is the bulk viscosity and 
\begin{equation}
		\sigma^{\mu \nu} = \Delta^{\alpha \mu} \Delta^{\beta \nu} \Bigg( \frac{\nabla_{\alpha} u_{\beta} + \nabla_{\beta} u_{\alpha}}{2} - 
			\frac{g_{\alpha \beta}}{3} \nabla_{\gamma} u^{\gamma} \Bigg),
\end{equation}
\begin{equation}
		\Delta^{\mu \nu} = u^{\mu} u^{\nu} + g^{\mu \nu}.
\end{equation}

	The simplest flow used to model heavy ion collisions is the so-called Bjorken flow, which possesses boost invariance in the longitudinal
	$z$ direction (the velocity profile is such that a Lorentz boost in the $z$ direction leaves the flow unchanged), rotational invariance 
	in the transverse plane (the $\phi$ direction),
	and translational invariance in the transverse plane.  What makes this flow so simple is that by the change of coordinates from $(t , x , 
	r ,\phi)$ to $(\tau , \eta , r , \phi)$ defined by
\begin{equation}
		t = \tau \cosh \eta \qquad z = \tau \sinh \eta
\end{equation}
	the four-velocity profile reduces to $u^{\mu} = (1 , 0 , 0 , 0)$ with a metric 
\begin{equation}
	ds^{2} = - d \tau^2 + \tau^2 d \eta^2 + dr^2 + r^2 d \phi^2.
\end{equation}
	
	In the past few years another analytical flow has been found that can be used to model heavy ion collisions.  Gubser \cite{Gubser 1},
	\cite{Gubser 2} exchanged the translational invariance in the transverse plane associated with Bjorken flow for conformal invariance.  
	This allows the fluid to expand radially at the expense of fixing the trace of the energy-momentum tensor to be zero.  Having a 
	radially-expanding analytical fluid flow opens up the possibility of investigating more realistic linear perturbations than were
	permitted by Bjorken flow.  In what follows, we shall refer to this new flow as Gubser flow.
	
	There exists a coordinate system in which Gubser flow is also particularly simple.  First, one can rescale the metric
\begin{equation}
		ds^2 = \tau^2 d \hat{s}^2.
\end{equation}
	Introducing the parameter $q$, which characterizes the transverse scale of the system (more precisely, $q^{-1}$ does this), and the new
	coordinates $\rho$ and $\theta$ defined by
\begin{align}
		\sinh \rho = - \frac{1 - q^2 \tau^2 + q^2 r^2}{2 q \tau},
\label{eq: Gubser coords 1} \\
		\tan \theta = \frac{2 q r}{1 + q^2 \tau^2 - q^2 r^2},
\label{eq: Gubser coords 2}
\end{align}
	one can complete the transformation.  The rescaled metric in the coordinates $(\rho , \eta , \theta , \phi)$ is given by
\begin{equation}
		d \hat{s}^2 = -d \rho^2 + d \eta^2 + \cosh^2 \! \rho \, (d \theta^2 + \sin^2 \theta \, d \phi^2),
\end{equation}
	and the fluid flow is simply $\hat{u}^{\mu} = (1 , 0 , 0 , 0)$. Note that because the Gubser flow is conformal, $\zeta = 0$, $\ve = 3p$,
	and $\ve \propto T^{4}$.
	
	In what follows we first examine rapidity independent perturbations of the Gubser flow. We begin by discussing first order perturbations
	to the Gubser flow in general, and then state the specific setup that we have examined. We then present our two approaches to the problem. 
	We state our results for this analysis, and compare them with previous work along these lines. After this semi-analytic approach, we 
	examine similar perturbations numerically in full nonlinear hydrodynamics.  The power spectra we obtain are analyzed, specifically for 
	the stability of the first minimum and for the dependence of the hydrodynamic response on $\eta / s$.  These results are stated and 
	the results of the two approaches are compared in a concluding section.

\section{$\eta$--independent perturbations of the Gubser flow}
\label{sec: methodology}

	We begin by outlining the general framework of first order perturbations to the Gubser flow.  As derived in \cite{Gubser 1}, the
	background temperature in the rescaled frame is given by
\begin{equation}
		\hat{T}_{b} = \Bigg( \frac{\hat{T}_{0}}{(\cosh \rho)^{2/3}} + \frac{h \sinh^{3} \rho}{9 (\cosh \rho)^{2/3}} \, 
		{}_{2}F_{1} \left( \frac{3}{2} , \frac{7}{6} ; \frac{5}{2} ; -\sinh^{2} \rho \right) \Bigg),
\label{eq: bg T}
\end{equation}
	where $h = \eta / T^{3}$, $\hat{T} = \tau \, 11^{1/4} \, T$, ${}_{2}F_{1}$ is a hypergeometric function, and here and in everything that
	follows, $q = (4.3 \, \text{fm})^{-1}$.  This value was used by Gubser \cite{Gubser 1} \cite{Gubser 2} as the value that best matched 
	experimental results at RHIC.  The parameter $\hat{T}_{0}$ characterizes the initial temperature of the flow.  The perturbation is implemented by
	varying the rescaled temperature and four-velocity by a small amount:
\begin{eqnarray}
		\hat{T} &=& \hat{T}_{b} (1 + \delta), \\
		\hat{u}_{\mu} &=& \hat{u}_{0 \, \mu} + \hat{u}_{1 \mu},
\end{eqnarray}
	where $\hat{u}_{0 \, \mu} = (-1 , 0 , 0 , 0)$ is the covariant components of the background four-velocity and
\begin{eqnarray}
		\hat{u}_{1 \, \mu} &=& (0 , 0 , u_{\theta}(\rho , \theta , \phi) , u_{\phi}(\rho , \theta , \phi) ), \\
		\delta &=& \delta(\rho , \theta , \phi).
\end{eqnarray}
	As detailed in \cite{Gubser 2} we can further split the spacial variables from the timelike one by projecting the scalar $\delta$ onto 
	the basis of spherical harmonics and the two nonzero components of the $\hat{u}_{1 \, \mu}$ onto the derivatives of the spherical
	harmonics using the decomposition
\begin{eqnarray}
		\delta &=& \sum_{L , M} c_{L M} \, \delta_{L}(\rho) Y_{L M}(\theta , \phi),
\label{eq: to match} \\
		\hat{u}_{1 \, i} &=& \sum_{L , M} c_{L M} \, v_{L}(\rho) \partial_{i} Y_{L M}(\theta , \phi),		
\end{eqnarray}
	with $i \in \{ \theta , \phi \}$.  Here, the $c_{L M}$ are simply the coefficients in these expansions (the ambiguity in their definition can
	be removed by demanding that $\delta_{L} (\rho_{0}) = 1$ for some fixed $\rho_{0}$).  In our analysis, we include all $L \leq 25$. We will often
	drop the $L$ subscript and simply write $\delta$ and $v$ when there is an equation that holds for each $L$.  Note that we are ignoring the 
	so-called vector modes	by making this decomposition, for the most general basis for vectors on $S^{2}$ depending on the coordinates 
	$(\rho , \theta , \phi)$ will be of the form
\begin{equation}
		v_{i} ( \rho , \theta , \phi) = v_{s}(\rho) \partial_{i} S(\theta , \phi) + v_{v}(\rho) V_{i}(\theta , \phi),
\end{equation}
	where $S$ is a scalar function and the divergence of $V$ on $S^{2}$ vanishes \cite{Gubser 2}.  The equations for $\delta$ and $v$ can be 
	written in the form
\begin{equation}
		\frac{d \vec{w}}{d \rho} = - \mathbf{\Gamma} \cdot \vec{w},
\label{eq: matrixeq}
\end{equation}
	where $\vec{w} = (\delta , v)$ and the components of the $\mathbf{\Gamma}$ matrix are
\begin{eqnarray}
          \Gamma_{11} &=& \frac{h \tanh^{2} \rho}{3 \hat{T}_{b}}, \\
		\Gamma_{12} &=& \frac{L (L + 1)}{3 \hat{T}_{b} \cosh^{2} \rho} \left( h \tanh \rho - \hat{T}_{b} \right), \\
		\Gamma_{21} &=& \frac{2 h \tanh \rho}{h \tanh \rho - 2 \hat{T}_{b}} + 1, \\
		\Gamma_{22} &=& \frac{8 \hat{T}_{b}^{2} \tanh \rho + h \hat{T}_{b} \left( \frac{-4 ( 3 L (L + 1) - 10)}{\cosh^{2} \rho} - 16 \right) 
					+ 6 h^{2} \tanh^{3} \rho}{6 \hat{T}_{b} \left( h \tanh \rho - 2 \hat{T}_{b} \right)}.
\end{eqnarray}
	This matrix equation $\eqref{eq: matrixeq}$ can be reduced to a second order equation for $\delta$:
\begin{equation}
		\frac{d^{2} \delta}{d \rho^{2}} + \frac{d \delta}{d \rho} \left( \Gamma_{11} - \frac{1}{\Gamma_{12}}\frac{d \Gamma_{12}}{d \rho} 
			+ \Gamma_{22} \right) + \delta \left( \frac{d \Gamma_{11}}{d \rho} - \frac{\Gamma_{11}}{\Gamma_{12}}\frac{d \Gamma_{12}}
				{d \rho} + \Gamma_{11}\Gamma_{22} - \Gamma_{12}\Gamma_{21} \right) = 0.
\label{eq: fulleq}
\end{equation}
	All of the information about the perturbed system follows from the solutions of \eqref{eq: fulleq}.
	
	We have used  \eqref{eq: fulleq} to investigate the evolution of an initial Gaussian hotspot
\begin{equation}
		\delta(\rho_{0} , \theta , \phi) \propto \exp \Bigg( \! \! - \frac{\theta^{2} + \theta^{2}_{0} - 2 \theta \theta_{0} \cos (\phi - \phi_{0})}{2 s^{2}} \Bigg),
\end{equation}
	with the further assumption that there be no initial velocity perturbations.  
\begin{equation}
		\hat{u}_{i}(\rho_{0}) = 0, \quad i \in \{ \theta , \phi \}.
\end{equation}
	We have done this in two separate ways (method A and method B), which we will outline in the next two subsections.  The $r_{0}$ coordinate was 
	varied, keeping $\tau_{0} = 1 \text{ fm}$ fixed for the hotspot.  This fixed a value for $\rho_{0}$ by the coordinate relations 
	\eqref{eq: Gubser coords 1}, \eqref{eq: Gubser coords 2}. A standard Cooper-Frye isothermal freezeout was then used in order to compute the
	particle spectrum of pions:
\begin{equation}
		E \frac{d N}{d^3 p} = - \int d \Sigma_{\mu} \, p^{\mu} f \left( \frac{p^{\nu} u_{\nu}}{T} \right).
\label{eq: CoopFrye}
\end{equation}
	The distribution function $f$ was taken to be a simple Boltzmann distribution.  Modifications of the freezeout surface were taken into
	account by keeping terms in the exponent of the Boltzmann distribution to the first order in $\delta \tau (r , \phi)$, the modification 
	to the freezeout time. 	After obtaining the particle spectrum, we calculated two-particle correlations by multiplying two single particle
	distributions and averaging over the $\phi$ component of the initial perturbation
\begin{equation}
		\frac{d N}{d (\Delta \phi)} = \int \! \frac{d N}{d (\phi_{1} - \psi)} \frac{d N}{d (\phi_{2} - \psi)} d \psi,
\end{equation}
	with $\Delta \phi = \phi_{1} - \phi_{2}$.  The two-particle distributions were then Fourier decomposed in order to generate a power 
	spectrum; that is, the coefficients $|v_{m}|^{2}$ in the expansion
\begin{equation}
		\frac{d N}{d \Delta \phi} = \left\langle \frac{d N}{d \Delta \phi} \right\rangle 
			\left( 1 + 2 \sum_{m = 1}^{\infty} |v_{m}|^{2} \cos (m \Delta \phi) \right) .
\end{equation}	

\subsection{First approach: Standard initial value problem}
\label{subsec: IVP}
	
	The first method that we used to solve \eqref{eq: fulleq} was by using a standard numerical ordinary differential equation solver.  The initial 
	conditions were set for each of the $L$ equations by decomposing the initial hotspot \eqref{eq: IC} into spherical harmonics.  These components
	were then set to be the coefficients $c_{L M}$ in \eqref{eq: to match}, and the initial condition for each $\delta$ was then simply $\delta_{L} 
	(\rho_{0}) = 1$.  The second initial condition was fixed by the demand that there be no initial fluid flow.  This is the approach followed in
	work by Staig and Shuryak \cite{Staig 2}.
	
\subsection{Second approach: The Green's Function Method}
\label{subsec: GFmethod}

	There is a completely different way to approach the initial value problem described above.  Consider a flow in which a perturbation 
	``turns on'' at an initial time $\rho_{0}$.
\begin{equation}
		T^{\mu \nu} = T^{\mu \nu}_{0} + \theta (\rho - \rho_{0}) \delta T^{\mu \nu}.
\label{eq: IC}
\end{equation}
	Here $T^{\mu \nu}$ is the total energy-momentum tensor for the system.  This leads to hydrodynamic equations of the form
\begin{equation}
		\nabla_{\mu} T^{\mu \nu} = \delta T^{\mu \nu} \partial_{\mu} \theta (\rho - \rho_{0}),
\end{equation}
	We can thus view an initial value problem for the hydrodynamic equations as a problem of a \emph{sourced} fluid flow for $\delta$.  With this 
	application in mind, we shall now outline how we construct the relevant Green's functions for this problem.
	
	Observe that the $\Gamma$s in \eqref{eq: fulleq} can be viewed as functions of $\tanh \rho$ only.  This means that we can rewrite the 
	equation in terms of the new variable $x \equiv \tanh \rho$.  If we further expand \eqref{eq: fulleq} to first order in $h$, we arrive
	at
\begin{align}
		(1 - x^{2})^{2} \delta''(x) - \frac{2}{3} x (1 - x^{2}) \delta'(x) + \frac{1}{3} L (L + 1) (1 - x^{2}) \delta(x) + \nonumber
			\\ + \frac{h}{3 T_{0}} (1 - x^{2})^{2/3} \Bigg[ \left( x \left( \frac{2}{1 - x^{2}} \right) - 2 x L (L + 1) \right) \delta(x) 
			+ \nonumber
				\\ + 3 \left( L (L + 1) (1 - x^{2}) + 3 x^{2} - 1 \right) \delta'(x) \Bigg] = 0,
\label{eq: first order}
\end{align}
	Now observe that for $\rho \in (- \infty , \infty)$, we have $x \in (-1 , 1)$.  This suggests that perhaps the ansatz
\begin{equation}
	\delta(x) = \sum_{n = 1}^{\infty} a_{n} P_{n}(x),
\label{eq: ansatz}
\end{equation}
	with $P_{n}$ the $n$th Legendre polynomial and $a_{n}$ the $n$th unknown coefficient, will be useful. If we substitute \eqref{eq: 
	ansatz} into \eqref{eq: first order} and then apply the integral operator $\int_{-1}^{1} dx \, P_{m}(x)$ to both sides of the equation,
	we arrive at an algebraic relation between the unknown coefficients $a_{i}$.  Unfortunately, in the case $h \neq 0$ these relations 
	involve an infinite number of the $a_{i}$.  If, however, we limit our ansatz to a finite expansion
\begin{equation}
	\delta(x) = \sum_{n = 1}^{N} a_{n} P_{n}(x),
\label{eq: finite ansatz}
\end{equation}
	we arrive at a set of linear relations between the coefficients $\{ a_{n} : n \in [0 , N] \cap \N \}$.  In our analysis, we take $N = 50$, when 
	solving the linear equations, but we only use the first $25$ terms when calculating results.  We use a larger $N$ when solving the linear
	equations than when generating results because truncating the series introduces errors, which mainly affect the larger $n$ coefficients.  If we
	regard this set as a column vector $a$ and the set of matrix coefficients in the resulting linear equations $M$, then we have reduced our
	original differential equation \eqref{eq: fulleq} to the linear equations 
\begin{equation}
		M a = 0.
\label{eq: matrix eqn}
\end{equation}

	For each $L$ in \eqref{eq: fulleq}, there should be two linearly independent solutions.  We can find them by considering the two 
	``initial conditions'' 
\begin{equation}
		(a_{0} , a_{1}) = (1 , 0) \quad \text{or} \quad (a_{0} , a_{1}) = (0 , 1).
\end{equation}
	In the inviscid case, these correspond to even and odd solutions respectively.  These initial conditions can be implemented as follows.  
	If we split the column vector $a$ as
\begin{equation}
		a = (a_{ic} \, | \, \overline{a})
\end{equation}
	and the matrix $M$ as
\begin{equation}
		M = (M_{1} \, | \, M_{2}),
\end{equation}
	where $a_{ic}$ is one of the two initial conditions, $\overline{a}$ are the $(N - 2)$ remaining coefficients, $M_{1}$ is a $(N - 2)
	\times 2$	matrix and $M_{2}$ is a square $(N - 2) \times (N - 2)$ matrix.  We then see that \eqref{eq: matrix eqn} can be rewritten as
	$M_{1} a_{ic} + M_2 \overline{a} = 0$, which allows one to solve for the unknown coefficients
\begin{equation}
		\overline{a} = (-M_{2}^{-1} M_{1}) a_{ic}.
\end{equation}

	We now address the question of the source.  Suppose we have a sourced hydrodynamic system with evolution equation 
\begin{equation}
		\nabla_{\mu} T^{\mu \nu} = J^{\nu}.
\end{equation}
	How does this source $J^{\nu}$ propagate through to the equation for $\delta$ \eqref{eq: fulleq}?  Recall that $\delta$ was defined by
\begin{equation}
		\hat{T} = \hat{T}_{b} (1 + \delta),
\end{equation}
	We can preform a similar decomposition to this source term as was done with the $\delta / v$ decomposition; that is
\begin{equation}
		 J^{\nu} = \sum_{L M} (J^{\rho}_{L} \, Y_{L M} , J^{A}_{L} \, \partial^{i} Y_{L M}), \quad i \in \{ \theta , \phi \}.
\end{equation}
	Using this decomposition, one finds that the source term on the right side of \eqref{eq: fulleq} is
\begin{equation}
		\left( \Delta_{1}' - \frac{\Gamma_{1 2}'}{\Gamma_{1 2}} \, \Delta_{1} + \Gamma_{2 2} \, \Delta_{1} \right) J^{\rho}_{L} -
			\Delta_{2} \, \Gamma_{1 2} \, J^{A}_{L} + \Delta_{1} (J^{\rho}_{L})',
\end{equation}
	where
\begin{equation}
		\Delta = \left( \frac{1}{4 \hat{T}_{b}^{4}} , \frac{3}{2 \hat{T}_{b}^{3} (2 \hat{T}_{b} - h \tanh \rho)} \right).
\end{equation}

	Using this source and the method described above, one can construct solutions of sourced hydrodynamic equations in the $(\rho, \eta , 
	\theta, \phi)$ coordinate system.  We have applied this method to the problem outlined above, and our results are described in the following 
	sections.  This method is also interesting in its own right and can be used to analyze two point functions directly, as has been investigated
	by Springer and Stephanov \cite{Springer}.  Letting $j(x)$ denote the source of the $\delta$ equation \eqref{eq: fulleq}, we can write a
	particular $L$ solution in the form
\begin{equation}
		\delta (x) = \delta_{\text{hom}}(x) + \int_{x_{0}}^{x} \! dx' G(x , x') j(x') 
\end{equation}
	with 
\begin{equation}
		G(x , x') = \frac{\Theta ( x - x' )}{W [f_{L} , f_{R}] (x')} \sum_{a} \sum_{b} (l_{a} r_{b} - l_{b} r_{a}) P_{a}(x') P_{b}(x).
\end{equation}
	Here, $\delta_{\text{hom}}$ is a solution to the homogeneous, unsourced equation, $\Theta$ is the Heaviside step function, and 
	$W[f_{L} , f_{R}]$ is the Wronskian of the solutions $f_{L}$ and $f_{R}$ that satisfy the left and right boundary conditions of interest. They
	have the expansions
\begin{equation}
		f_{L}(x) = \sum_{a} l_{a} P_{a}(x) \quad \text{and} \quad f_{R}(x) = \sum_{a} r_{a} P_{a}(x)
\end{equation}
	respectively.  Since we have constructed the actual Green's functions, we can investigate the true two point functions for a Dirac-$\delta$
	source using this method.

\section[results]{Results}

\begin{figure}[hb]
\raggedright
\includegraphics[width=0.97\textwidth]{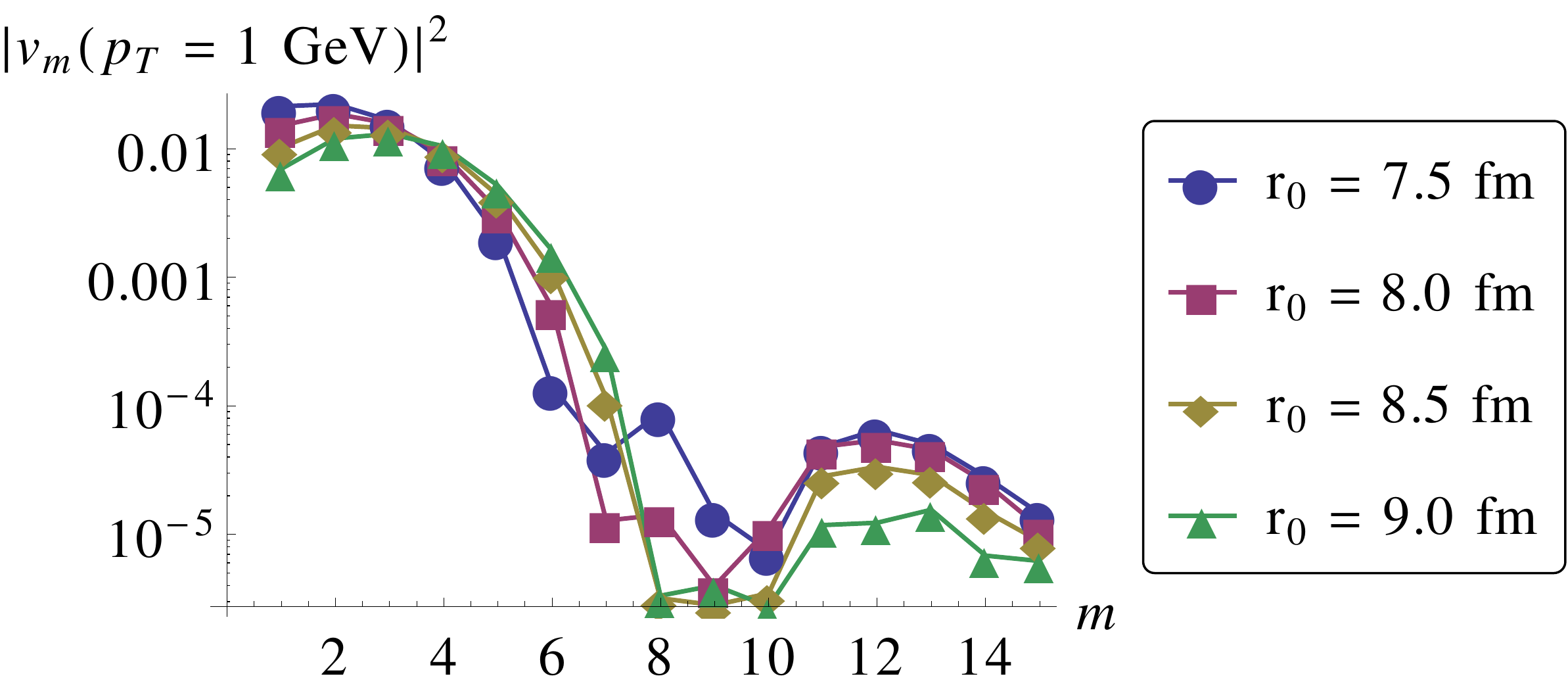}
\caption{Fixed $p_{T} = 1$ GeV differential power spectrum for $\eta = 0$. Numerical error is smaller than the symbol size.}
\label{fig: inviscid}
\end{figure}

	We have calculated the power spectrum for a first order perturbation using method B for $\hat{T}_{0}
	= 7.9$, which corresponds to an initial temperature of $T_{0} = 500$ MeV.  The sample inviscid plot is shown in Figure \ref{fig: inviscid}.
	The width of the Gaussian hotspot used in our analysis was $s = 0.2$, to make it significantly more narrow than the 
	background flow representing the colliding nuclei.  Viscous plots looked qualitatively similar and have not been included, for we have followed Staig and
	Shuryak \cite{Staig 2} in ignoring viscous corrections to the spectrum.

	In Figure \ref{fig: inviscid} we see that the position of the first minimum of the power spectrum is not stable to
	changes in the initial radial position of the perturbation $r_{0}$.  Increasing $r_{0}$ tends to shift the first minimum of the power spectrum
	to larger $m$ values.  We observe the minimum to vary by two or three harmonics, which is broadly consistent with
	Staig and Shuryak \cite{Staig 2}.  In what follows, however, we note a number of differences between their work and ours (see, in particular 
	the following two subsections \ref{subsec: compare SS} and \ref{subsec: v_n over epsilon_n}).
	It seems reasonable to conclude that the first minimum of the fixed $p_{T}$ differential power spectrum is not a 
	stable observable, even in linear hydrodynamics.  If one were to average over many events with multiple
	perturbations, there would not be a well-defined location of this first minimum.  Since real hydrodynamics is nonlinear, it seems
	doubtful that there will be a detectable first minimum to the fixed $p_{T}$ differential power spectrum in real heavy ion collision data.  
	
	Let us briefly digress in order to perform an error analysis of our method. There are two main sources of 
	numerical error: a Riemann sum was used in the integration 
	process and the freezeout surface was truncated at the edge in order to not need to access times $\rho < \rho_{0}$ (see the comparison
	subsection below for a discussion of this).  The Riemann sum error was estimated by performing sums with two different resolutions; and the
	error associated with the truncation of the freezeout surface was was estimated by varying where this truncation was made, calculating the
	changes to the power spectra, and then extrapolating these changes to the edge of the freezeout surface.  The total error was found to
	be $2 - 3 \%$, in the region plotted in Figure \ref{fig: inviscid}, with the larger errors associated to the
	minima of the power spectra. 

\begin{figure}[t]
\centering
\begin{subfigure}[l]{0.45\textwidth}
\includegraphics[width = \textwidth]{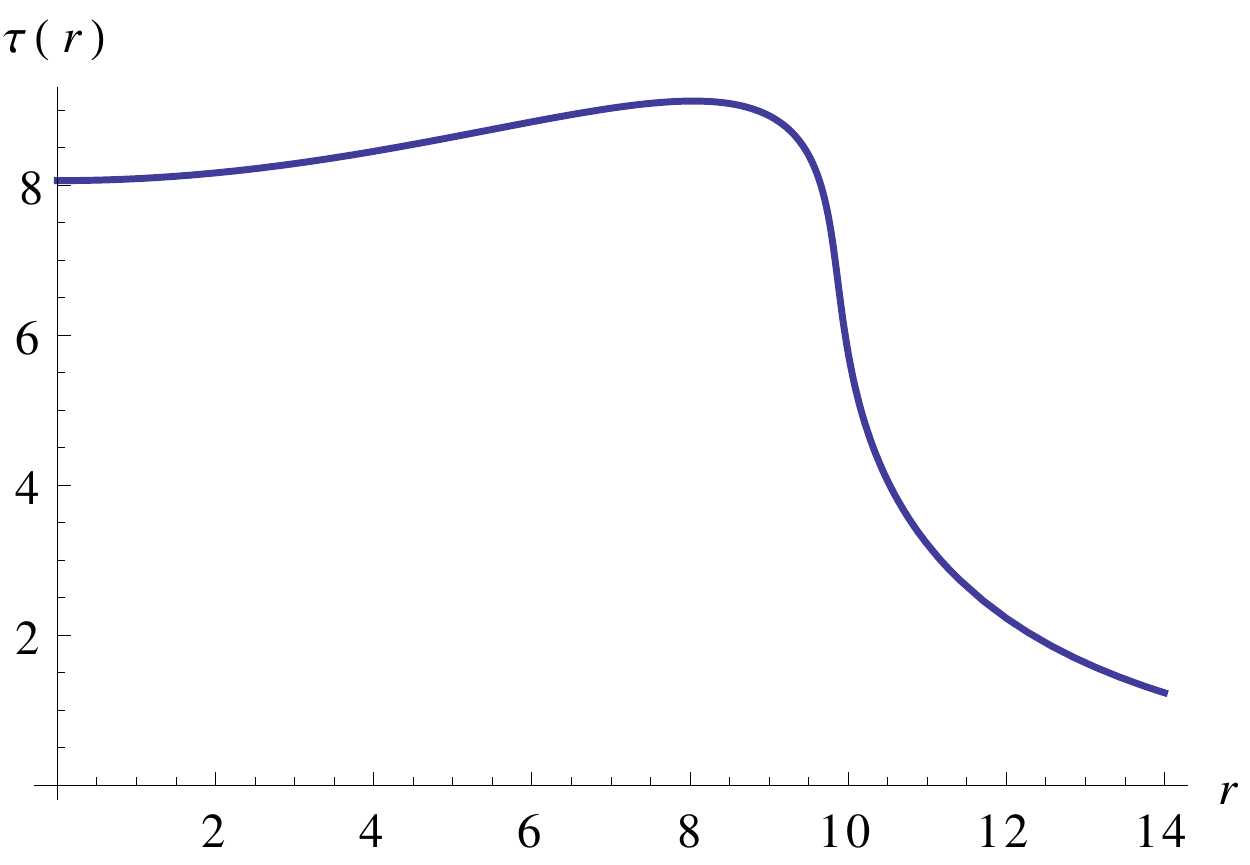}
\end{subfigure}
\qquad
\begin{subfigure}[r]{0.45\textwidth}
\includegraphics[width = \textwidth]{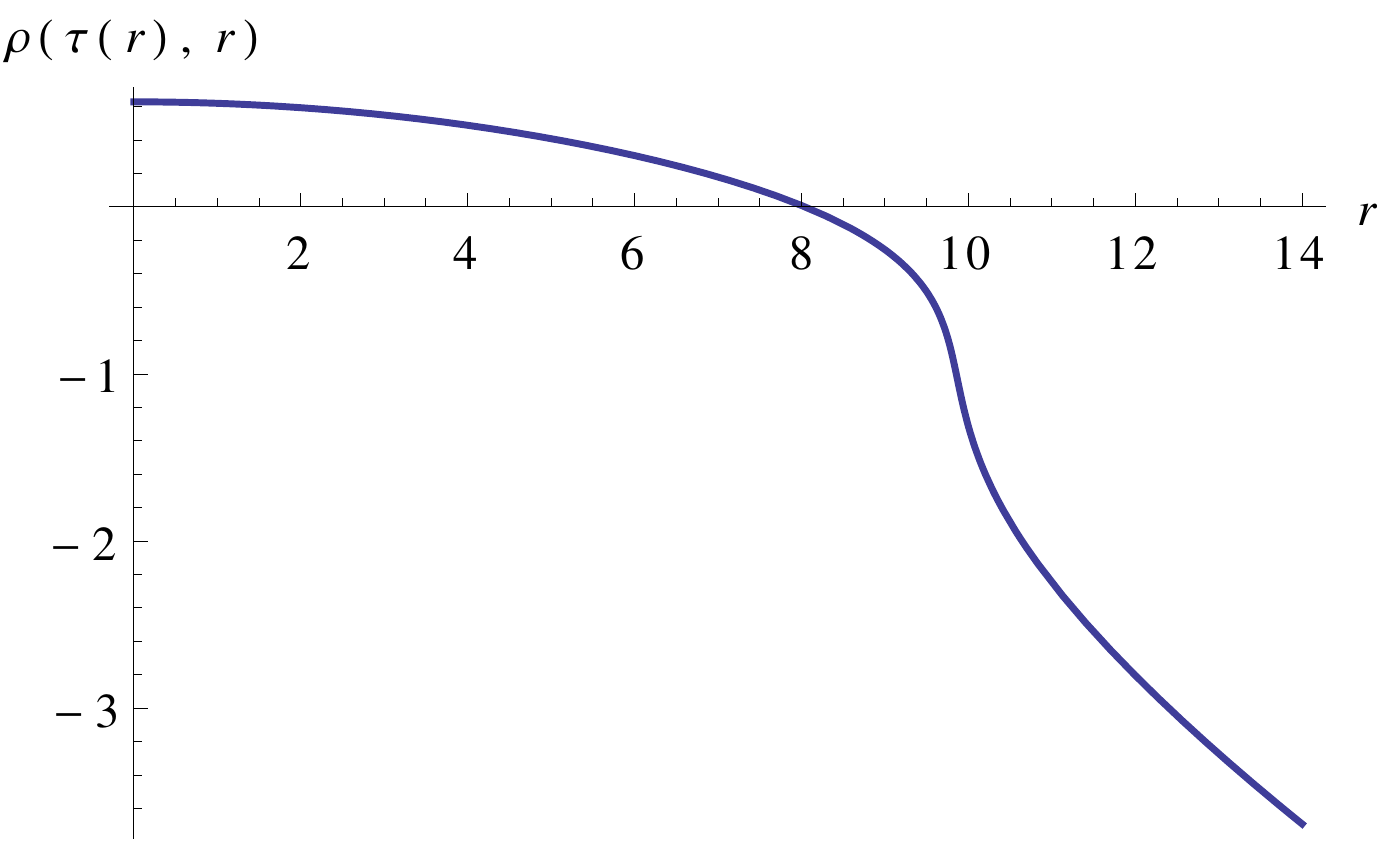}
\end{subfigure}
\caption{Left: a plot of the freezeout surface for $T_{0} = 630$ MeV.  Right: the $\rho$ values corresponding to this freezeout surface.}
\label{fig: rho naught}
\end{figure}

	We have also tried to identify potential systematic errors in our study and found one issue that needs to be taken into account. First, 
	recall that the Gubser timelike coordinate $\rho$ is not the same as the physical proper 
	time $\tau$.  From the coordinate transformations \eqref{eq: Gubser coords 1}, \eqref{eq: Gubser coords 2} one sees that the 
	coordinate $\rho$ depends on both of the physical coordinates $\tau$ and $r$.  This means that for a given freezeout surface there is a 
	restricted range of $\rho_{0}$ that one must choose from in order to avoid needing information about the flow \emph{before} the perturbation 
	is initialized (when calculating the Cooper-Frye freezeout integral \eqref{eq: CoopFrye}, for example). Figure 
	\ref{fig: rho naught} illustrates this effect.  
	On the left, the freezeout surface for $T_{0} = 630$ MeV is plotted.  On the right, the value of the coordinate $\rho$ is show for the
	points on the freezeout surface.  We see that if we wish to avoid freezing out for $\rho < \rho_{0}$ at the edge of the freezeout
	surface, we mush either choose a $\rho_{0}$ sufficiently early, or only integrate out to a certain $r$ in the Cooper-Frye formula
	\eqref{eq: CoopFrye}; that is, ignore parts of the freezeout surface.  Since we wish to fix the $\tau_{0}$ coordinate of our
	perturbation to be $1 \text{ fm}$, this leads to the necessary choice of ignoring the very edge of the freezeout surface.  In order 
	to keep this systematic error small, we chose to
	cut the freezeout surface between $r = 12.3 \text{ fm}$ and $r = 12.6 \text{ fm}$ in order to include as much of the features of the
	surface as possible.  This allowed us to choose the $r_{0}$ value of the perturbation to be anything larger than about $8 \text{ fm}$
	(or, for $T_{0} = 500$ MeV, about $7.5$ fm).  This is a rather large value, but bringing it closer to the center would have necessitated
	ignoring significant portions of the freezeout surface, and in turn would result in a large systematic error of our procedure.

\subsection{Remarks on differences from previous works}
\label{subsec: compare SS}

	Staig and Shuryak \cite{Staig 2} have analyzed the same problem that we have outlined above as an initial value problem.  In their
	paper, they found that the first minimum of the fixed $p_{T}$ differential power spectrum did not shift more than one harmonic as they 
	varied the $r_{0}$ position between $3$ fm and $5.5$ fm, and the overall shape of the power spectra remained consistent as the coordinate was 
	varied.  Despite the broadly consistent shape of the differential power spectra between our work and theirs, there are some subtle differences 
	in analysis.  For this reason, we highlight here the few differences between our analysis and that of Staig and Shuryak.
	
	Staig and Shuryak presumably evaluated the full Cooper-Frye integral \eqref{eq: CoopFrye} with a saddle point approximation, where the
	integrand was evaluated at the maximum of the background exponent. In our analysis, we found that the integrand in the Cooper-Frye 
	expression \eqref{eq: CoopFrye} was not sharply peaked for every value of $\phi_{P}$, the	momentum angular coordinate, so a full 
	calculation of the particle spectra was necessary.  We evaluated the integral in full, using only a Riemann sum in the $r$ coordinate;
	our numerical errors due to this sum were discussed above.
	
	We also used a smaller value of $\hat{T}_{0}$ than Staig and Shuryak.  They chose to use $\hat{T}_{0} = 10.1$, corresponding to an
	initial temperature of $T_{0} = 630$ MeV, in order to allow more time to evolve before freezeout.  This changes the
	detailed shape of the power spectra, and shifts the peaks and minima somewhat, but it does not change our qualitative conclusion that
	the first minimum is not stable under variations in the initial coordinate $r_{0}$ of the hotspot.

	Finally, there is no mention in Staig and Shuryak's work of the systematic error introduced related to the cutting of the freeze-out surface 
	(see our discussion above).  Note, however, that they used values of $r_{0}$ that necessitate cutting a large portion of the freezeout surface
	to avoid freezing out for $\rho < \rho_{0}$.


\subsection{Hydrodynamic response of the Gubser flow}
\label{subsec: v_n over epsilon_n}

\begin{figure}[ht]
\centering
\includegraphics[width = 0.97 \textwidth]{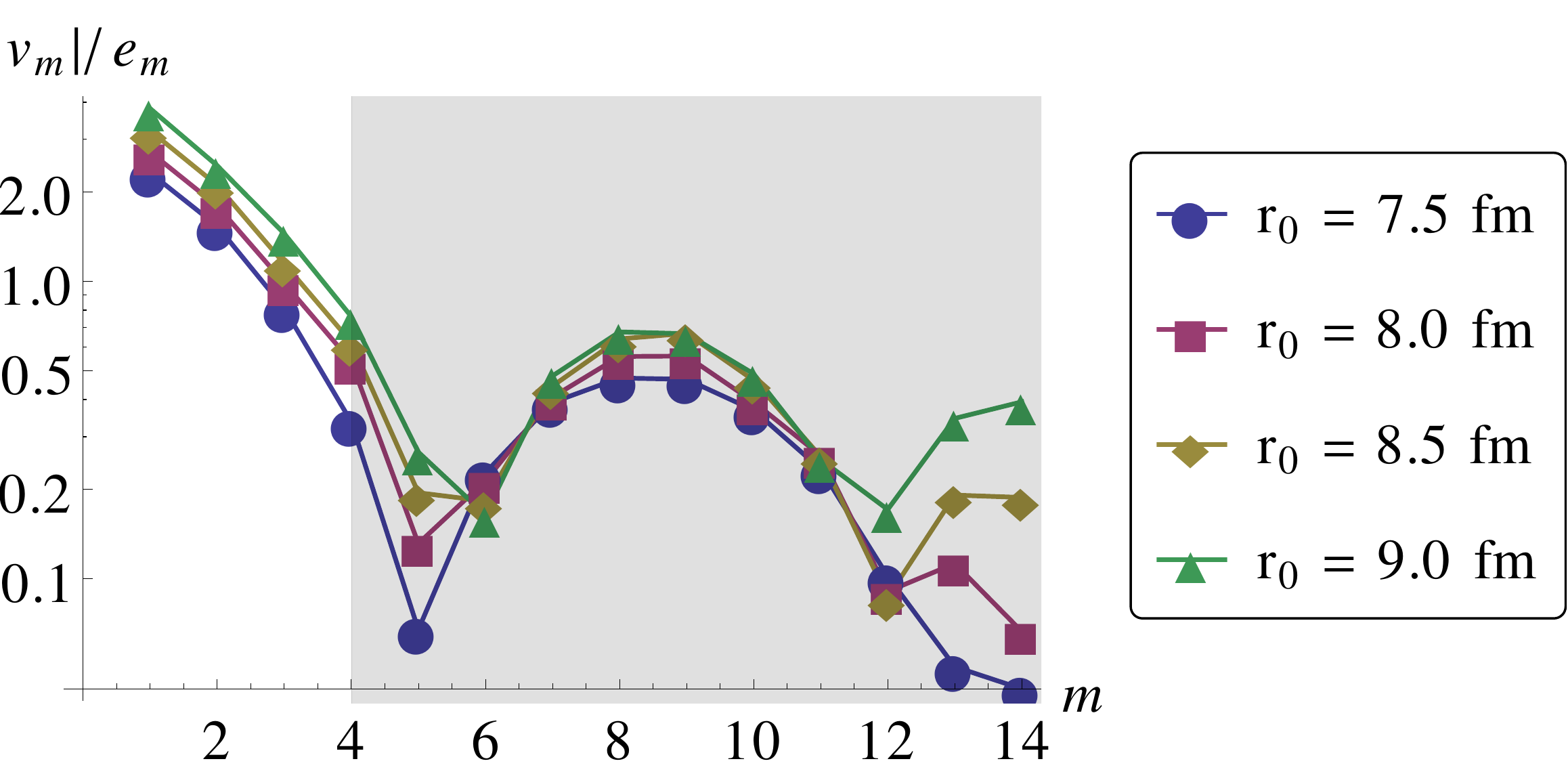}
\caption{Inviscid response. Numerical error is smaller than the symbol size. The shaded region $m > 3$ is cutoff-dependent.}
\label{fig: inviscid response}
\end{figure}

	In this section we study the linear hydrodynamic response to the perturbations by calculating the ratio $v_{m} / e_{m}$ where $e_{m}$ are the
	initial eccentricities.  Again, for the initial eccentricities, we have used the normalization
\begin{equation}
		e_{m} = \left| \frac{\int \! r^{m} \ve( x ) e^{i m \phi} d^{2} x}{\int \! r^{m} \ve( x ) d^{2} x} \right|,
\end{equation}
	where here $\ve \propto T^{4}$ is the initial energy density, and the integrals are over the transverse plane of the collision.  The
	hydrodynamic response of the system is frequently characterized by the ratio $v_{m} / e_{m}$.  We have calculated this ratio; in particular
	we have calculated it for integrated $v_{m}$ over the $p_{T}$ range $0 - 2$ GeV. This was performed with the weighting
\begin{equation}
		v_{m}^{(\text{integrated})} = \frac{\int d p_{T} \, p_{T} \, v_{m}( p_{T} ) \frac{d N}{d p_{T}}}{\int d p_{T} \, p_{T} \, 
			\frac{d N}{d p_{T}}},
\end{equation}
	where the integrals are over $0 - 2$ GeV.  Our results for the inviscid case are plotted in Figure \ref{fig: inviscid response}.  
	The temperature of the fluid and width of the perturbation are identical to what was used in the corresponding plots above.

	We note here that the $e_{m}$ are ill-defined for the Gubser flow, for $\ve_{0}(r)$, the background energy density, exhibits a power law 
	decay. From the background temperature profile \eqref{eq: bg T}, we find the leading order behavior
\begin{equation}
		\ve_{0}(r) \propto r^{-16 / 3} + \cdots.
\end{equation}
	Since the background energy density is independent of the azimuthal angle $\phi$ and the perturbation is a Gaussian, we have that the 
	numerators of the expression for $e_{m}$ are finite for every $m$. On the other hand, the denominators 
	diverge for large enough $m$, for they are given by
\begin{equation}
		\int^{R} \! r^{m} ( r^{- 16 / 3} + \cdots ) r \, d r \propto R^{(m - 10 / 3)} + \cdots \to  \infty
\end{equation}
	if $m > 3$. In order to remedy this, we have imposed a cutoff on the volume by only integrating to a finite $r_{\text{max}}$.  This 
	cutoff, however, artificially causes the response to scale with $r_{\text{max}}$ as
\begin{equation}
		\frac{v_{m}}{e_{m}} \propto \left( r_{\text{max}} \right) ^{m - 10 / 3} + \cdots
\end{equation}
	for $m > 3$. This consideration must be held in mind when examining the hydrodynamical response of the Gubser flow.  To aid with this, we have
	shaded the cutoff-dependent region in Figure \ref{fig: inviscid response}.

\section{Calculating the $v_{m}$ using numerical hydrodynamics}
\label{sec: numerical hydro}
 
	Using the Gubser flow to model the energy density profile of the quark gluon plasma after a heavy ion collision was useful because it 
	provided an analytical flow on which to do linear perturbations.  However, it was not without its drawbacks.  The Gubser flow was 
	conformal, whereas the flow in a real heavy ion collision is not, and there were limitations on where the original perturbation could be 
	placed.  The latter difficulty necessitated discarding small pieces of the freezeout surface in the Cooper-Frye integral \eqref{eq: 
	CoopFrye}.  Finally, real viscous hydrodynamics is inherently nonlinear, whereas the Gubser analysis above only involved linear 
	perturbations.  With these shortcomings in mind, we have conducted analysis of the $v_{m}$ coefficients using numerical hydrodynamics.  

\begin{figure}[ht]
\centering
\includegraphics[width = 0.97 \textwidth]{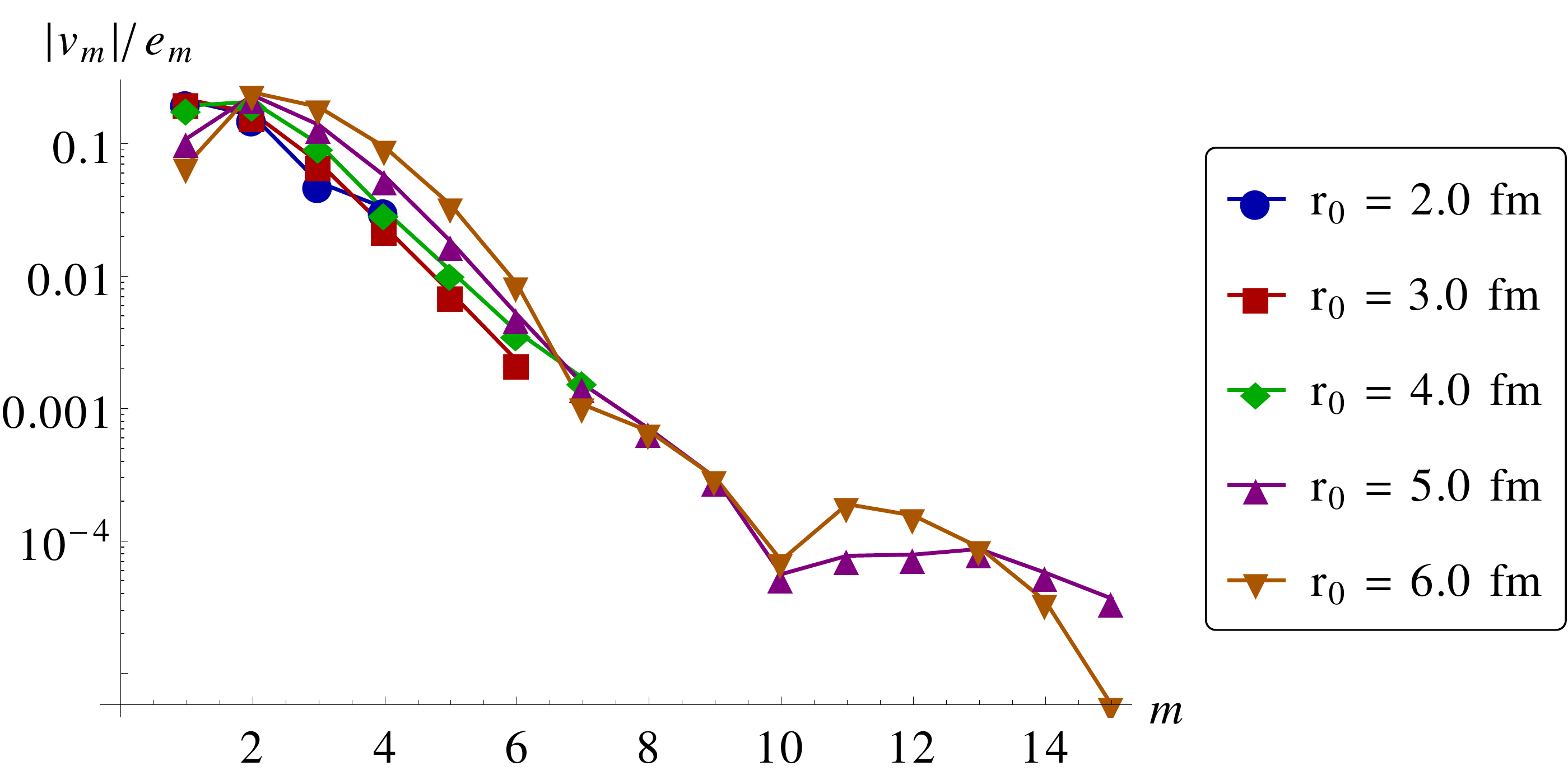}
\caption{Response in non-linear hydrodynamics. Here $\eta / s = 0.08$. Errors are no larger than the symbol size.}
\label{fig: hydroCodeResponse}
\end{figure}

	The specific code used was the Causal Viscous Hydro Code for Non-Central Heavy Ion Collisions version 0.5.2 by Luzum and Romatschke \cite{
	Code 1}.  On top of the central collision background energy density used in the code, a Gaussian of width $s = 0.5$ fm and height $0.125 
	\text{ GeV}^{4}$ was added, again, so that the perturbation would be localized relative to the background energy density.
	Figure \ref{fig: hydroCodeResponse} shows the integrated response over the $p_{T}$ range $0 - 4$ GeV of a system with $\eta / s = 0.08$ as 
	the initial radial position of the perturbation is varied between $2 - 6$ fm. The curves representing perturbations with smaller $r_{0}$ are 
	truncated at smaller $m$.  This is because smaller $r_{0}$ leads to smaller $v_{m}$ and $e_{m}$, which the code could not accurately calculate 
	with feasible lattice spacings.  We did, however, run codes for multiple lattice spacings in order to get a sense of the errors.  We have 
	only plotted those points that have errors smaller than the symbol size on the plots.

	From Figure \ref{fig: hydroCodeResponse} there does not appear to be a well-defined minimum until at least $m = 10$, except perhaps for 
	the perturbation farthest from the symmetry axis. After averaging over many events with many initial perturbation locations, one would 
	not expect to find a minimum for $m < 10$, which makes it doubtful that experiments would be able to detect one. 

	With this observation in mind, we turn to examining the proposed form for the response  
\begin{equation}
		\ln \left( \frac{v_{m}}{e_{m}} \right) \propto - \frac{4}{3 R T} m^{2} \frac{\eta}{s},
\label{eq: response form}
\end{equation}
	to see whether numerical hydrodynamics supports this simple dependence on $\eta / s$.
\begin{figure}[ht]
\centering
\includegraphics[width = 0.90 \textwidth]{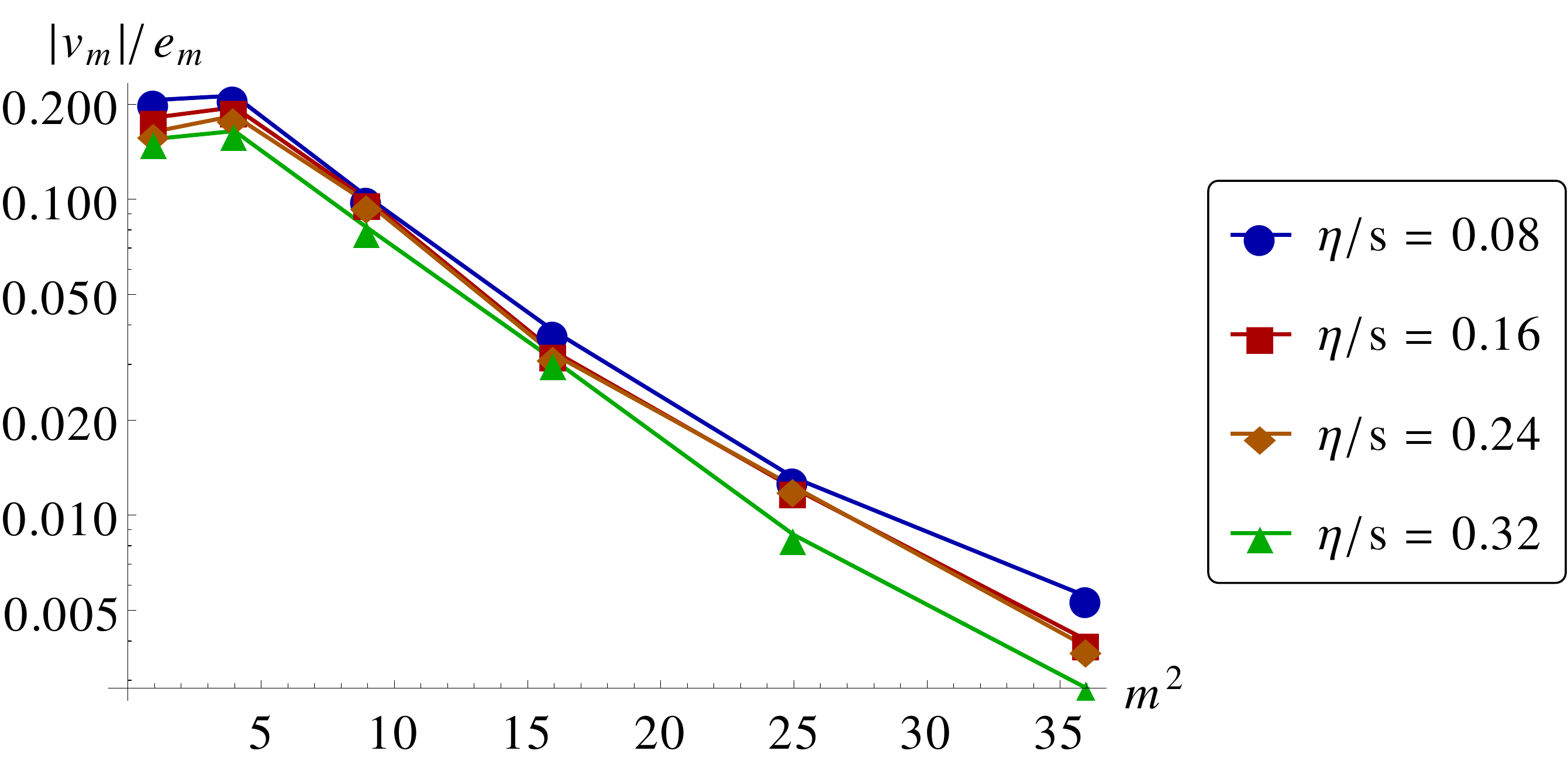}
\caption{Hydrodynamic response for different values of $\eta / s$. Note that this is a log plot and that the horizontal axis is $m^{2}$.  
	 Here, $r_{0} = 4$ fm. Errors are no larger than the symbol size.}
\label{fig: varyETAOS}
\end{figure}
\begin{figure}[ht]
\centering
\includegraphics[width = 0.90 \textwidth]{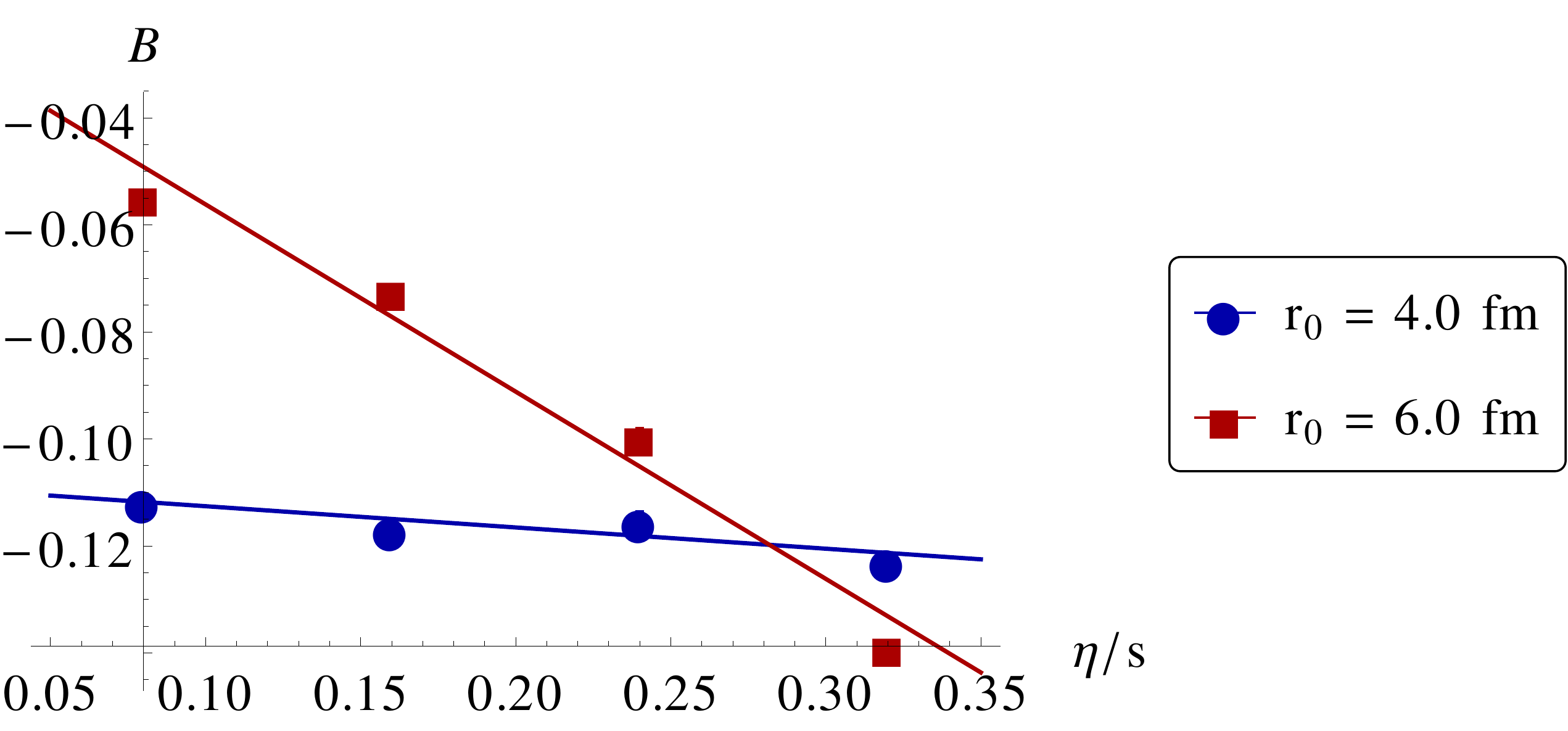}
\caption{Fit of the coefficients of $m^{2}$ in quadratic fits of $\ln \! \left( v_{m} / e_{m} \right)$ (cf. \eqref{eq: response form}). Each individual 
	fit of $\ln \! \left( v_{m} / e_{m} \right)$ includes the values $m = 2$ to $m = 6$. Errors are smaller than the symbol size.}
\label{fig: fit ETAOS}
\end{figure}
	Figure \ref{fig: varyETAOS} shows plots of $v_{m} / e_{m}$ for $\eta / s = 0.08, \, 0.16, \, 0.24, \text{ and } 0.32$ for perturbations 
	centered at $4$ fm from the beamline. As expected, we see that $v_{2}$ decreases as $\eta / s$ increases.  We note here, however, that
	there is not a noticeable change in slope as $\eta / s$ is varied, already placing strain on the proposed linear 
	dependence.  For each value of $\eta / s$ we have fit $\ln \left( v_{m} / e_{m} \right)$ to the form 
\begin{equation}
		A + B m^{2},
\end{equation}
	the proposed form of the response \eqref{eq: response form}.  Since the coefficients $B$ are proposed to depend linearly on $\eta / s$, we 
	show in Figure \ref{fig: fit ETAOS} a linear fit to these ($\eta / s$, $B$) values for the $r_{0} = 4$ fm and the $r_{0} = 6$ fm cases.  
	For each value of viscosity over entropy density, we plot the $B$ 
	value for the finest lattice spacing with error bars equal in size to the difference between the $B$ values for the finest and second
	finest lattice spacings (most of which are too small to be seen on the plot).

	The $A + B m^{2}$ form was found to fit the logarithm of the response for each viscosity over entropy ratio well, but we see from Figure 
	\ref{fig: fit ETAOS} that as the position of the perturbation is varied, the slope of the best fit line changes dramatically.  It would thus 
	seem from this work that after many events are averaged over one would no longer expect to see this dependence of the response \eqref{eq: response form} 
	on $m$ and $\eta / s$.

\section{Conclusions}

	We have used semi-analytical and numerical methods to investigate the hydrodynamical response of a heavy ion collision system with initial 
	state perturbations.  Semi-analytically, using the Gubser flow, we found that the first minimum of the response was sensitive to the location 
	of the initial-state perturbation.  
	Moreover, this minimum was located in a region where the initial state eccentricities were ill-defined.
	Numerically, we found that there did not appear to be a well-defined first minimum in the region $m < 10$. Taking these results together, we 
	conclude that in the region $m < 10$, which includes the region accessible by experiment, there should not occur a sharp minimum of the 
	response. Furthermore, our investigations do not 
	support the dependence of the measured response on $m$ and $\eta / s$ as
\begin{equation}
		\ln \! \left( \frac{v_{m}}{e_{m}} \right) \propto - m^{2} \frac{\eta}{s},
\end{equation}
	for the slope in the linear fits of $\ln \left( v_{m} / e_{m} \right)$ vs. $\eta / s$ varies with the location of the initial perturbation.
	In addition, for some locations, the linear dependence is not even observed.
	We thus find that many of the features thought to contain 
	information about the medium will be washed out when averaging over many event locations, or at the very least they will be modified 
	substantially.	
	
	We have also described a method for calculating the Green's functions for the $\eta$-independent, linearized Gubser flow to 
	first order in $h$.  This would allow one to examine the true two-point functions of this system at central rapidity.  The method could also
	in principle be extended to remove the central rapidity restriction.

\section{Acknowledgements}

	This work was supported by the Sloan Foundation, Award No. BR2012-038, and the DOE, Award No. DE-SC0008132. We thank Pilar Staig and Edward 
	Shuryak for an illuminating correspondence and J. Nagle for fruitful discussions.


\begin{thebibliography}{99}
%
	\bibitem{Sorensen} P. Sorensen, arXiv:nucl-ex/1201.0784v1.

	\bibitem{Thermal QGP} P. Braun-Munzinger, K. Redlich, J. Stachel, arXiv:nucl-th/0304013v1.

	\bibitem{Shuryak} E. Shuryak,  Phys. Rev. \textbf{C80}, 054908 (2009).

	\bibitem{vn paper} A. P. Mishra, R. K. Mohapatra, P. S. Saumia, A. M. Srivastava, Phys. Rev. \textbf{C77}, 064902 (2008).

	\bibitem{Staig 1} P. Staig, E. Shuryak, Phys. Rev. \textbf{C84}, 034908 (2011). 

	\bibitem{Staig 2} P. Staig, E. Shuryak, Phys. Rev. \textbf{C84}, 044912 (2011).

	\bibitem{Lacey} R. A. Lacey, A. Taranenko, J. Jia, D. Reynolds, N. N. Ajitanand, \emph{et al.} Phys. Rev. Lett. \textbf{112}, 082302 (2014).

	\bibitem{Gubser 1}  S. S. Gubser, Phys. Rev. \textbf{D82}, 085027 (2010).

	\bibitem{Gubser 2}  S. S. Gubser, A. Yarom, Nucl. Phys. \textbf{B846}, 469-511 (2011).

	\bibitem{Springer}  T. Springer, M. Stephanov, Nucl. Phys \textbf{A904-905}, 1027c, (2013),  arXiv:1210.5179v1 [nucl-th].

	\bibitem{Code 1} M. Luzum and P. Romatschke, Phys. Rev. \textbf{C78}, 034915 (2008)

%
\end{thebibliography}
\end{document}